\documentclass[%
aps, prl,
floatfix,
preprintnumbers,
reprint,
superscriptaddress,
amsmath, amssymb,
longbibliography,
]{revtex4-1}

\usepackage[utf8]{inputenc}
\usepackage{graphicx}
\usepackage{bm}
\usepackage[breaklinks=true]{hyperref}
\hypersetup{bookmarksnumbered, pdfpagemode=UseOutlines, 
	pdfauthor={K.\ S.\ Das}, 
	pdftitle={Anisotropic Hanle lineshape via anisotropic magnetothermoelectric phenomena}, 
	pdfdisplaydoctitle, 
	colorlinks=true, citecolor=blue, filecolor=blue, linkcolor=blue, urlcolor=blue}

\begin{document}
	
	\title{Anisotropic Hanle lineshape via anisotropic magnetothermoelectric phenomena}
	
	\author{K.\ S.\ \surname{Das}}
	\email[e-mail: ]{K.S.Das@rug.nl}
	\affiliation{Zernike Institute for Advanced Materials, University of Groningen, 9747 AG Groningen, The Netherlands}
	\author{F.\ K.\ Dejene}
	\affiliation{Max Planck Institute of Microstructure Physics, D-06120 Halle, Germany}
	\author{B.\ J.\ \surname{van Wees}}
	\affiliation{Zernike Institute for Advanced Materials, University of Groningen, 9747 AG Groningen, The Netherlands}
	\author{I.\ J.\ Vera-Marun}
	\email[e-mail: ]{ivan.veramarun@manchester.ac.uk}
	\affiliation{Zernike Institute for Advanced Materials, University of Groningen, 9747 AG Groningen, The Netherlands}
	\affiliation{School of Physics and Astronomy, University of Manchester, Manchester M13 9PL, UK}
	
	
	\begin{abstract}
		We observe anisotropic Hanle lineshape with unequal in-plane and out-of-plane non-local signals for spin precession measurements carried out on lateral metallic spin valves with transparent interfaces. The conventional interpretation for this anisotropy corresponds to unequal spin relaxation times for in-plane and out-of-plane spin orientations as for the case of 2D materials like graphene, but it is unexpected in a polycrystalline metallic channel. Systematic measurements as a function of temperature and channel length, combined with both analytical and numerical thermoelectric transport models, demonstrate that the anisotropy in the Hanle lineshape is magneto-thermal in origin, caused by the anisotropic modulation of the Peltier and Seebeck coefficients of the ferromagnetic electrodes. Our results call for the consideration of such magnetothermoelectric effects in the study of anisotropic spin relaxation.
	\end{abstract}
	
	\pacs{72.25.-b, 85.75.Ff, 85.80.-b, 72.15.Jf}
	\keywords{spin polarized transport, spin precession, thermoelectrics, anisotropic magnetotransport}
	
	
	\maketitle
	
	
	Electrical spin injection and detection in non-local lateral spin valves have been used extensively to study pure spin currents in non-magnetic (NM) materials \cite{johnson_interfacial_1985,jedema_electrical_2001,jedema_electrical_2002-1,jedema_spin_2003,valenzuela_spin-polarized_2004,kimura_large_2007,tombros_electronic_2007,kimura_temperature_2008}. Hanle measurements allow the manipulation of the spin accumulation in the NM via a perpendicular magnetic field, which induces spin precession as the carriers diffuse along the NM channel. From these experiments, we can extract the spin transport parameters of the channel, like the spin relaxation length and time, and hence get an insight about the nature of spin-orbit interaction (SOI) causing spin relaxation. This is particularly relevant for 2D materials like graphene, where the SOI acting along the in-plane and out-of-the plane directions can differ and lead to anisotropic spin relaxation, manifested by different signals for the in-plane and out-of-plane spin configurations in the Hanle experiments \cite{tombros_anisotropic_2008,guimaraes_controlling_2014}. In contrast, for polycrystalline films, spin relaxation is expected to be isotropic \cite{zutic_spintronics:_2004}.

	In this work we use metallic non-local spin valves (NLSVs), with aluminium (Al) as the NM material, to study spin precession as a function of temperature. Permalloy (Ni$_{80}$Fe$_{20}$, Py) has been used as the ferromagnetic (FM) electrodes to inject a spin-polarized current into Al across a transparent interface and to non-locally detect the non-equilibrium spin accumulation in Al at a distance $L$ from the injector. This model system with transparent FM/NM interfaces has been thoroughly studied via spin valve measurements. But curiously, corresponding spin precession studies in such systems are scarce. Only recently a few groups have demonstrated spin precession in NLSVs with transparent FM/NM interfaces \cite{idzuchi_effect_2014,villamor_effect_2015}, with the NM channel being either silver or copper. More importantly, these few experiments have been done only at low temperatures ($T\leq 10$~K), with no reports on Hanle measurements at room temperature for transparent FM/NM interfaces.

	We demonstrate, through non-local spin precession experiments on Py/Al NLSVs with transparent interfaces, an anomalous Hanle lineshape for $T> 150$~K, in which the in-plane and out-of-plane spin signals are unequal. This anisotropic Hanle lineshape generally indicates different spin relaxation rates for spins aligned parallel and perpendicular to the plane of the NM channel \cite{tombros_anisotropic_2008,guimaraes_controlling_2014}. However, anisotropic spin relaxation in a polycrystalline metallic film has not been observed in the literature and is unexpected, especially being stronger at higher temperatures. Such a temperature dependence of the anisotropy is indicative of a thermoelectric origin. With the help of analytical and numerical thermoelectric transport models, we ascribe the anisotropy in the Hanle measurements to a change in the baseline resistance \cite{bakker_interplay_2010} due to the anisotropy in the Seebeck and Peltier coefficients of the FM. The results evidence how an apparent anisotropic spin precession can develop in an isotropic NM channel, via the coexistence of spin and heat currents and spin-orbit coupling in the FM.

\begin{figure}[tbp]
		\includegraphics*[angle=0, trim=0mm 0mm 0mm 0mm, width=85mm]{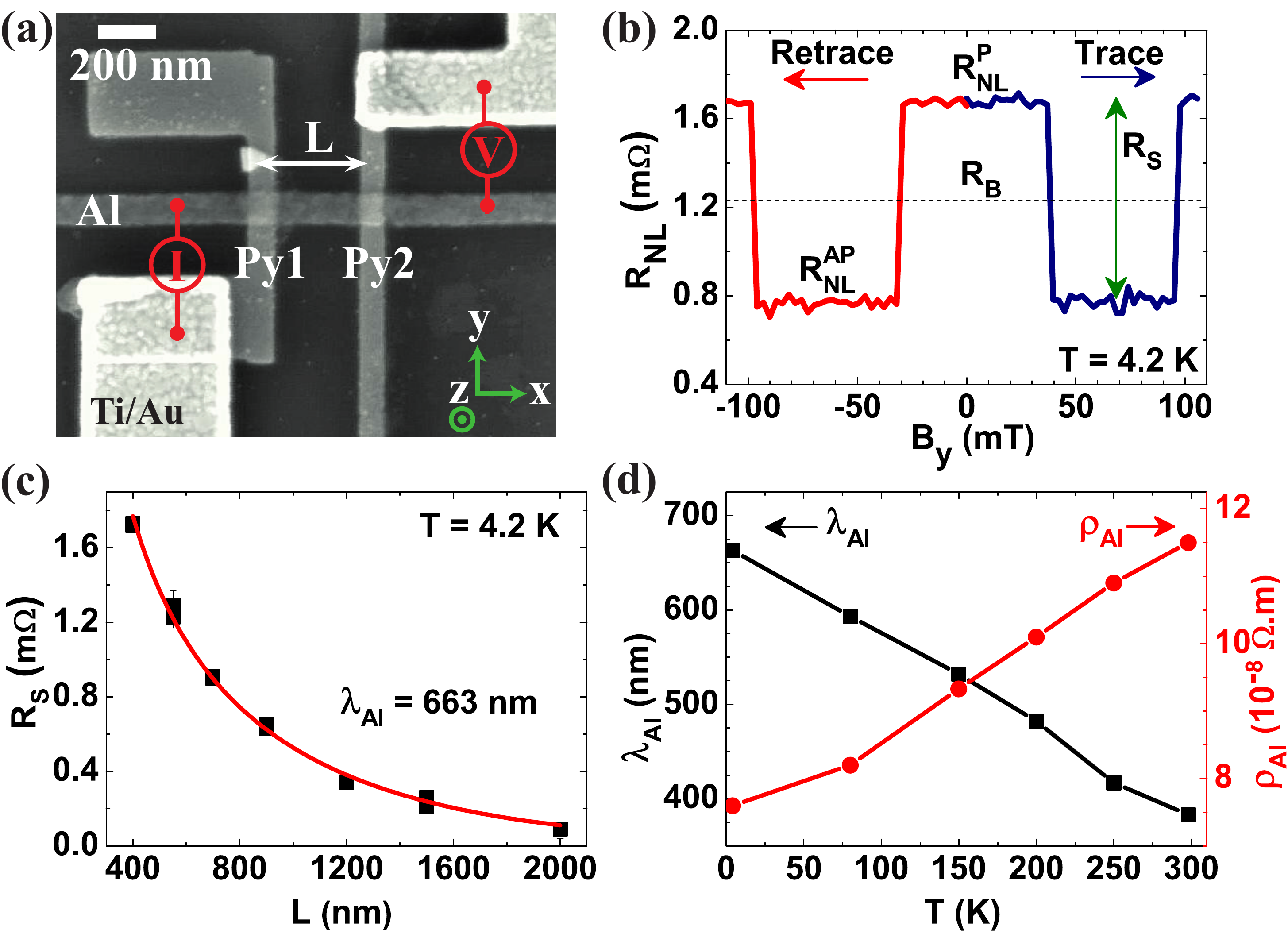}
		\caption{
			\label{fig:NLSV}
			\textbf{(a)} An SEM image of a representative NLSV along with the electrical connections for spin-valve and Hanle measurements. Py1 and Py2 act as spin injector and detector, respectively, separated by a distance $L$. \textbf{(b)} Spin-valve measurement on a device with $L$ = 700~nm at $T$ = 4.2~K. The parallel ($R_{\text{NL}}^{\text{P}}$) and anti-parallel ($R_{\text{NL}}^{\text{AP}}$) states are shown along with the baseline resistance ($R_{\text{B}}$) and the spin accumulation signal ($R_{\text{S}}$). \textbf{(c)} Dependence of $R_{\text{S}}$ on $L$, used to extract the spin relaxation length in Al ($\lambda_{\text{Al}}$), by fitting the data (black squares) with a spin diffusion model (red line) as described in the text. The error bars correspond to the noise (standard deviation) in the spin-valve curves when quantifying $R_{\text{NL}}^{\text{P}}$ and $R_{\text{NL}}^{\text{AP}}$ signals. \textbf{(d)} Temperature dependence of $\lambda_{\text{Al}}$ and the resistivity of the Al channel ($\rho_{\text{Al}}$).
		}
	\end{figure}

	
	Py/Al NLSVs with transparent interfaces (interface resistance $<10^{-15}$~$\Omega.\text{m}^2$) and varying injector-detector separations ($L$) were prepared on top of a 300~nm thick SiO$_2$ layer on a Si substrate. The device preparation is described in detail in the supplementary material \cite{see_supplementary_material_see_????} and follows Refs.~\cite{kimura_large_2007,villamor_effect_2015,bakker_interplay_2010}. Fig.~\ref{fig:NLSV}(a) shows an SEM image of a representative NLSV along with the electrical connections for spin-valve and Hanle measurements. A low frequency alternating current ($I = 400$ $\mu$A) was applied between the injector (Py1) and the left end of the Al channel. The first harmonic response of the corresponding non-local signal ($R_{\text{NL}} = V_{\text{NL}}/I$) was measured between the detector (Py2) and the right end of the Al channel by standard lock-in technique.

	The NLSVs were first characterized via spin-valve measurements as shown in Fig.~\ref{fig:NLSV}(b). An external magnetic field ($B_{\text{y}}$) was swept along the main axis of the FMs to orient their magnetization in either parallel (P) or anti-parallel (AP) configurations, corresponding to distinct levels $R_{\text{NL}}^{\text{P}}$ and $R_{\text{NL}}^{\text{AP}}$ in the non-local response. From these measurements we extracted the spin accumulation signal in the Al channel, $R_{\text{S}} = R_{\text{NL}}^{\text{P}}-R_{\text{NL}}^{\text{AP}}$, and the baseline resistance, $R_{\text{B}} = (R_{\text{NL}}^{\text{P}}+R_{\text{NL}}^{\text{AP}})/2$ (which later will be used to interpret the spin precession measurements). The spin accumulation created at the injector junction decays exponentially in the Al channel with a characteristic spin relaxation length, $\lambda_{\text{Al}}$. Fig.~\ref{fig:NLSV}(c) shows the dependence of $R_{\text{S}}$ on the injector-detector separation ($L$), from which $\lambda_{\text{Al}}$ can be extracted using the standard spin diffusion formalism for transparent contacts \cite{takahashi_spin_2003,valet_theory_1993, schmidt_fundamental_2000}. We extracted $\lambda_{\text{Al}}$ to be 663~nm at 4.2~K and 383~nm at 300~K. A systematic study of the temperature dependence of $\lambda_{\text{Al}}$ revealed its monotonic decrease with increasing $T$, with an opposite behaviour for the resistivity of the channel ($\rho_{\text{Al}}$), as shown in Fig.~\ref{fig:NLSV}(d). These results are consistent with Elliott-Yafet spin relaxation mechanism dominated by electron-phonon interaction in bulk metal \cite{zutic_spintronics:_2004,kimura_temperature_2008,mihajlovic_surface_2010}, in which the spin relaxation length is proportional to the electron mean free path.

	\begin{figure}[tbp]
		\includegraphics*[angle=0, trim=0mm 0mm 0mm 0mm, width=85mm]{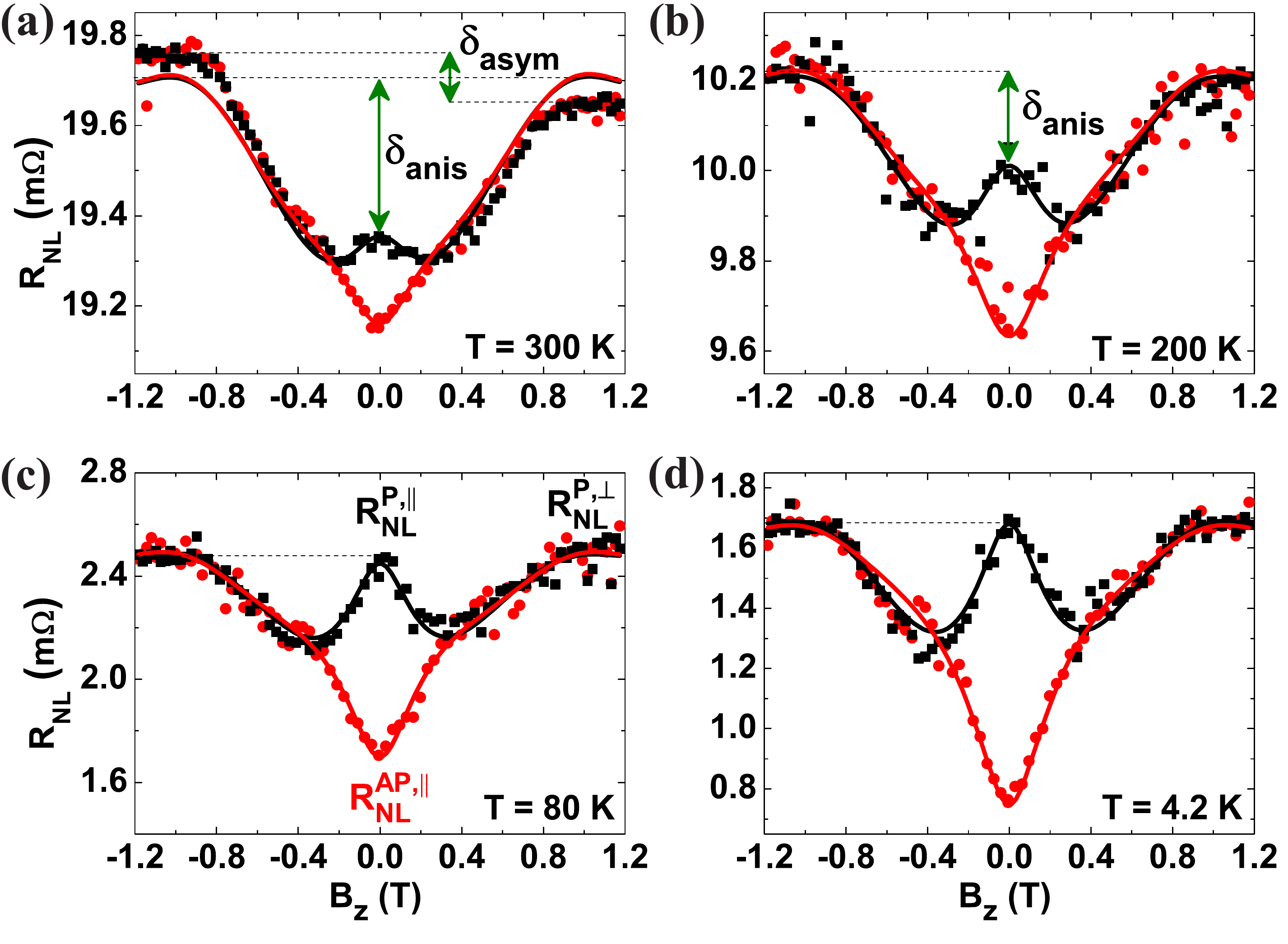}
		\caption{
			\label{fig:Hanle}
			Hanle measurements in a NLSV with $L$ = 700~nm at different temperatures: \textbf{(a)} $T$ = 300~K, \textbf{(b)} $T$ = 200~K, \textbf{(c)} $T$ = 80~K and \textbf{(d)} $T$ = 4.2~K. The initial magnetic configuration of the FM contacts at $B_{\text{z}}=0$ is in-plane and either parallel ($R_{\text{NL}}^{\text{P},\parallel}$, black squares) or anti-parallel ($R_{\text{NL}}^{\text{AP},\parallel}$, red circles), whereas for $|B_{\text{z}}|>$ 0.9~T it is out-of-plane and parallel ($R_{\text{NL}}^{\text{P},\perp}$) . The anisotropy ($\delta_{\text{anis}}$) in the non-local signal ($R_{\text{NL}}$) between spins oriented in-plane (y) and out-of-plane (z) is observed at 300~K and 200~K, but it is absent at 80~K and 4.2~K. The solid lines are fits to the Hanle data (see text).}
	\end{figure}

	Next, we perform Hanle spin precession measurements, in which a perpendicular magnetic field ($B_{\text{z}}$) induces the spins injected into the Al channel to precess at a Larmor frequency $\omega_{\text{L}}=g\mu_{\text{B}}B_{\text{z}}/\hbar$, where $g\approx2$ is the $g$-factor in Al, $\mu_{\text{B}}$ is the Bohr magneton and $\hbar$ is the reduced Planck constant. As shown in Fig.~\ref{fig:Hanle}(a-d), Hanle measurements can be performed with the magnetizations of the FMs initially aligned in-plane (at $B_{\text{z}} = 0$) and set either parallel (P) or anti-parallel (AP) with respect to each other. The Larmor precession and the resulting spin dephasing, lead to a decrease (increase) in the signal $R_{\text{NL}}$ with increasing $|B_{\text{z}}|$ for the P (AP) configuration, eventually intersecting the AP (P) curve for an average spin rotation of $\pi/2$. After the intersection of the P and AP curves, they bend upwards with increasing $|B_{\text{z}}|$ and finally saturate for $|B_{\text{z}}|\geq 0.9$~T. This happens because the magnetization of Py starts to rotate out-of-plane and finally aligns with $B_{\text{z}}$ for $|B_{\text{z}}|\geq 0.9$~T. The rotation of Py's magnetization with $B_{\text{z}}$ can be checked from the anisotropic magnetoresistance (AMR) measurements of the Py wire, described in the supplementary material \cite{see_supplementary_material_see_????} and follows Refs.~\cite{rijks_semiclassical_1995,jedema_electrical_2002-1}. Thus, for $|B_{\text{z}}|\geq 0.9$~T, the spins are injected (and detected) in the out-of-plane (z) direction and there should be no precession caused by $B_{\text{z}}$. For isotropic spin relaxation and parallel orientation of the magnetizations, the signal $R_{\text{NL}}^{\text{P},\parallel}$ for spins injected in-plane at $B_{\text{z}}=0$ should be equal to the signal $R_{\text{NL}}^{\text{P},\perp}$ when spins are injected out-of-plane at $|B_{\text{z}}|\geq 0.9$~T. We indeed observe that $R_{\text{NL}}^{\text{P},\parallel}=R_{\text{NL}}^{\text{P},\perp}$ for the Hanle data at 80~K and 4.2~K (Fig.~\ref{fig:Hanle}(c) and (d)). These Hanle data were fitted with an analytical expression obtained by solving the Bloch equation considering spin precession, diffusion and relaxation for transparent contacts \cite{fukuma_giant_2011,villamor_effect_2015} and taking into account the out-of-plane rotation of the Py magnetization \cite{jedema_electrical_2002-1}. From the fitting, we obtained $\lambda_{\text{Al}}$ to be 688~nm at 4.2~K and 544~nm at 80~K, which are comparable to the values obtained from the spin-valve measurements (Fig.~\ref{fig:NLSV}(d)).

	At higher temperatures ($T\geq150$~K), we notice a significant difference between $R_{\text{NL}}^{\text{P},\parallel}$ and $R_{\text{NL}}^{\text{P},\perp}$, leading to anisotropic Hanle lineshapes as shown in Figs.~\ref{fig:Hanle}(a) and (b). Such Hanle lineshapes have been hitherto associated with anisotropic spin relaxation \cite{tombros_anisotropic_2008, guimaraes_controlling_2014}, in which the NM channel has different spin relaxation times for the in-plane and out-of-plane spin directions. For isotropic and polycrystalline metallic films, as is the case for our 50~nm thick Al channel, the transverse and longitudinal spin relaxation times are expected to be equal \cite{zutic_spintronics:_2004}. Moreover, by increasing the temperature we expect any anisotropy to decrease due to the thermal disorder in the system. Hence we rule out anisotropic spin relaxation in our system and investigate other causes for the observed Hanle lineshapes. Further checks were performed to rule out: (i) the role of interfacial roughness and magnetic impurities by probing the presence of inverted Hanle response \cite{dash_spin_2011,mihajlovic_magnetic-field_2011} in the spin-valve measurements at high in-plane fields ($B_{\text{y}}$) and (ii) non-linear effects by measuring higher harmonics and at different current densities. For details of these further checks, see the supplementary material \cite{see_supplementary_material_see_????}. 
	
	We quantify the anisotropy in the Hanle measurements by the parameter $\delta_{\text{anis}}=R_{\text{NL}}^{\text{P},\perp}-R_{\text{NL}}^{\text{P},\parallel}$, as shown in Figs.~\ref{fig:Hanle}(a) and (b). We note that concurrent to this anisotropy we also observe a smaller asymmetry with the sign of $B_{\text{z}}$, $\delta_{\text{asym}}=R_{\text{NL}}^{\text{P},\perp}(B_{\text{z}}<-0.9$~T$)-R_{\text{NL}}^{\text{P},\perp}(B_{\text{z}}>0.9$~T$)$, as shown in Fig.~\ref{fig:Hanle}(a). Since $\delta_{\text{asym}} \ll \delta_{\text{anis}}$ we focus the discussion below on the anisotropy ($\delta_{\text{anis}}$).

\begin{figure}[tbp]
		\includegraphics*[angle=0, trim=0mm 0mm 0mm 0mm, width=85mm]{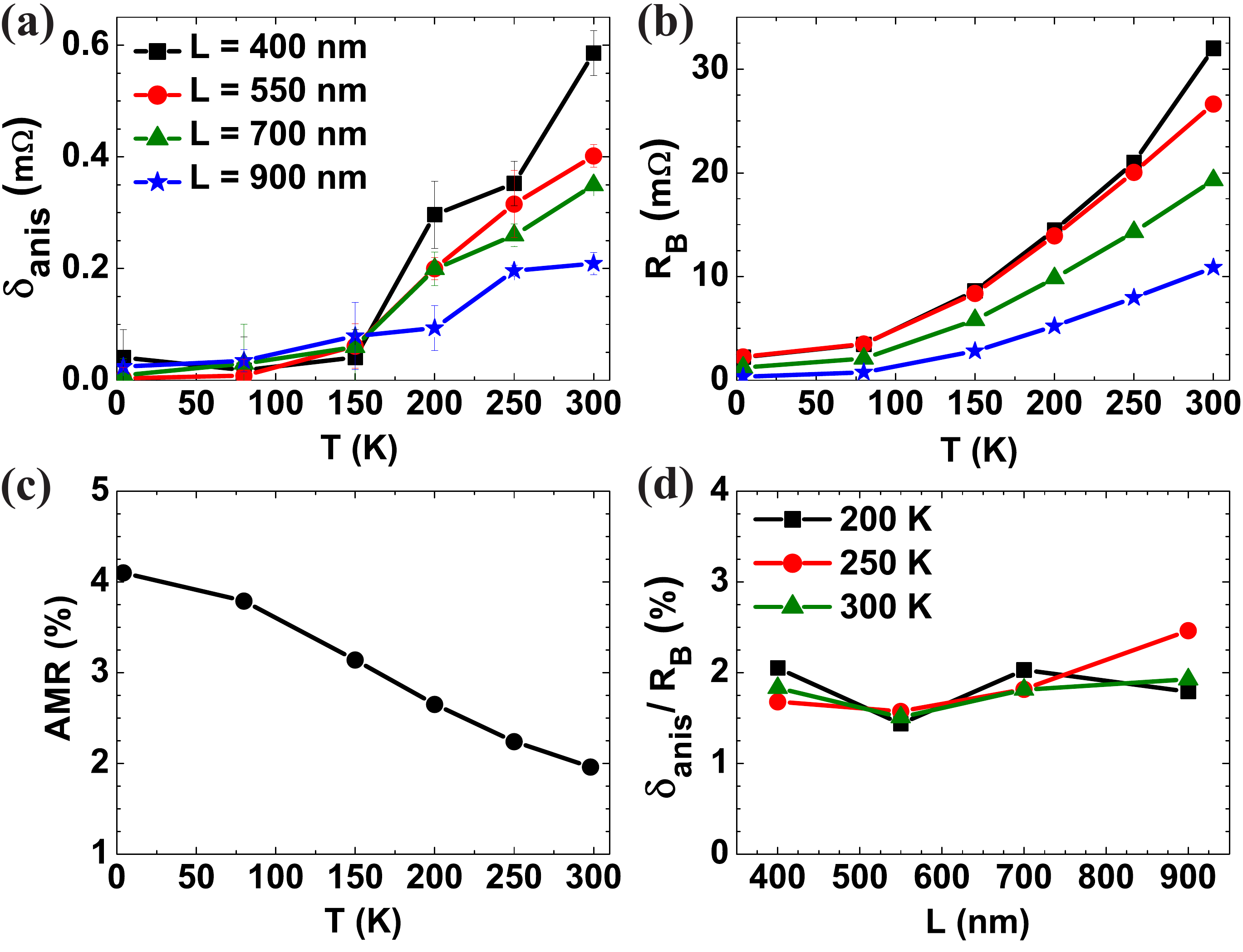}
		\caption{
			\label{fig:Anisotropy}
			Temperature ($T$) dependence of: \textbf{(a)} The anisotropy ($\delta_{\text{anis}}$) extracted from Hanle measurements for different channel lengths ($L$), \textbf{(b)} The baseline resistance ($R_{\text{B}}$) extracted from spin-valve measurements, and \textbf{(c)} Anisotropic magnetoresistance (AMR) of Py. \textbf{(d)} A constant ratio $\delta_{\text{anis}}/R_{\text{B}}\approx2\%$ is observed, independent of $L$ and $T$.
		}
	\end{figure}

	A marked non-linear increase with temperature is observed on both the anisotropy $\delta_{\text{anis}}$ (extracted from Hanle measurements), and the baseline resistance $R_{\text{B}}$ (obtained from spin-valve measurements) in the measurements summarized in Fig.~\ref{fig:Anisotropy}(a-b). We interpret these observations as an indication for a common thermal origin for both effects. 
	Note that these trends are inconsistent with an effect purely related to spin currents, as $\lambda_{\text{Al}}$ decreases at higher $T$ (Fig.~\ref{fig:NLSV}(d)). Furthermore, the trends are also inconsistent with the trivial effect of AMR on local charge currents, because the AMR has also an opposite trend with temperature (Fig.~\ref{fig:Anisotropy}(c)). 
	We remark that the origin of $R_{\text{B}}$ in NLSVs has been identified as thermoelectric in nature \cite{bakker_interplay_2010}. It is driven by the interplay of Peltier cooling and heating at the injector junction, in which a charge current across the junction results in a temperature difference, and the Seebeck effect at the detector junction, which acts as a nanoscale thermocouple to electrically detect the non-local heat currents. Here, we hypothesize that the anisotropy $\delta_{\text{anis}}$ is also thermoelectric in nature, in particular given the striking observation of an almost constant ratio $\delta_{\text{anis}}/R_{\text{B}}\approx 2\%$ independent of $L$ and $T$, as shown in Fig.~\ref{fig:Anisotropy}(d).

	To further understand the origin of the anisotropy $\delta_{\text{anis}}$, we must note that $|B_{\text{z}}|$ modulates the magnetization direction of Py, which together with Al forms thermoelectric junctions. Similarly as the electrical resistance of Py gets modulated due to AMR, we consider here a modulation in the Seebeck ($S$) and Peltier ($\Pi$) coefficients as a function of the angle between the magnetization and the heat current, i.e.\ anisotropic thermoelectric transport due to spin-orbit  interaction in the FM \cite{wegrowe_anisotropic_2006,slachter_anomalous_2011,mitdank_enhanced_2012,kimling_anisotropic_2013}. To test this hypothesis, we develop a thermoelectric model to estimate $R_{\text{B}}$ in our NLSVs, and relate its corresponding magnetothermoelectric effect to $\delta_{\text{anis}}$.
	
\begin{figure}[tbp]
		\includegraphics*[angle=0, trim=0mm 0mm 0mm 0mm, width=85mm]{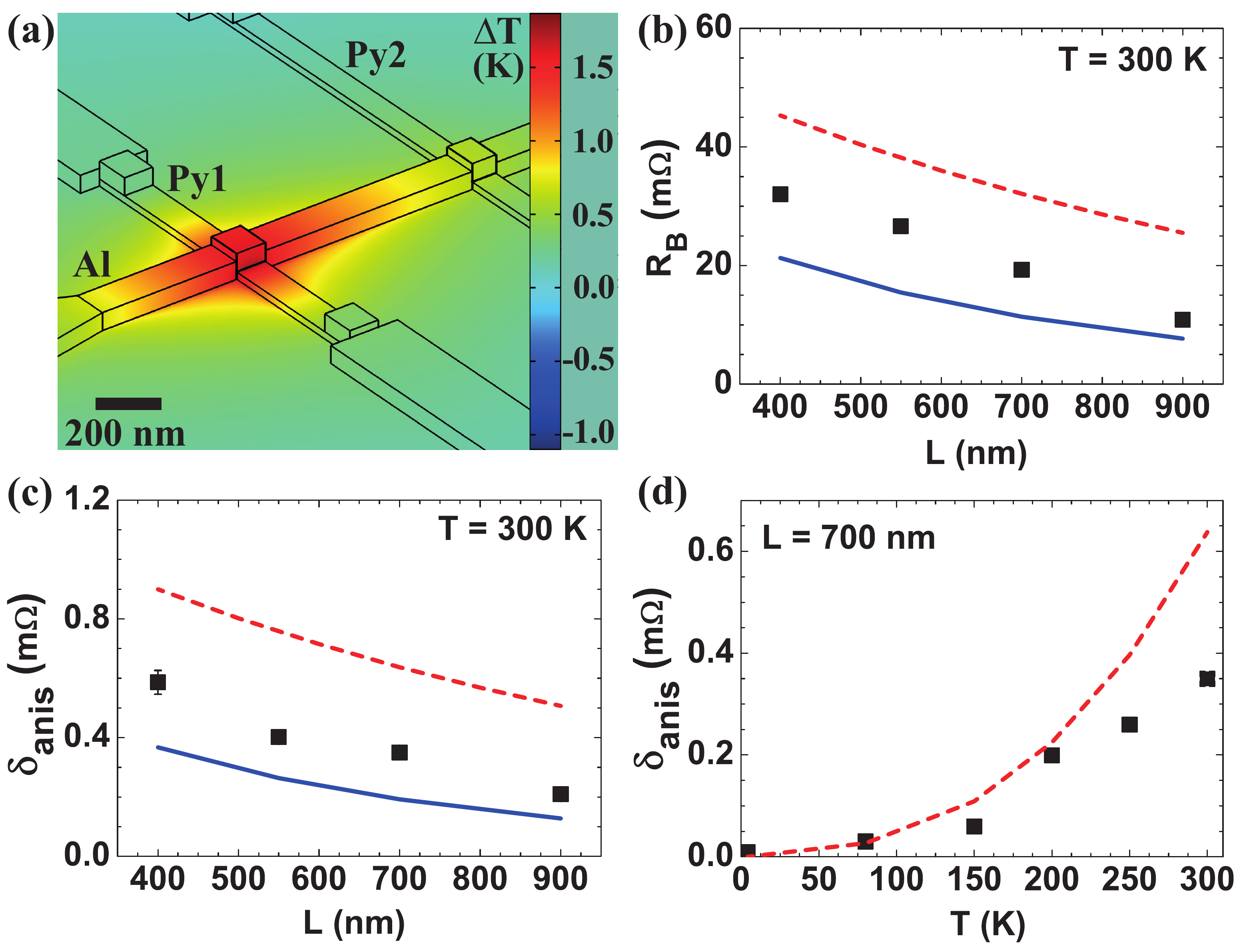}
		\caption{
			\label{fig:Simulation}
			\textbf{(a)} The temperature difference ($\Delta T$) in the device area, simulated by 3-dimensional finite element modelling (3D-FEM), is shown as a colour map. Comparison between the measured data (black squares), the analytical model (red dashed lines) and 3D-FEM (blue solid lines) is presented for the dependence of the: \textbf{(b)} Baseline resistance ($R_{\text{B}}$) and \textbf{(c)} Anisotropy ($\delta_{\text{anis}}$) on the channel length ($L$) at 300~K. \textbf{(d)} Temperature dependence of $\delta_{\text{anis}}$, obtained experimentally and through the analytical model, for a fixed channel length of 700~nm.
		}
	\end{figure}

	The Peltier effect at the injector junction results in a temperature difference ($\Delta T$), with respect to the reference temperature ($T$), equal to 
	\begin{equation}
		\Delta T = \dot{Q}R_{\text{th}} = (\Pi_{\text{Al}}-\Pi_{\text{Py}})IR_{\text{th}},
		\label{eq:PeltierTemperature}
	\end{equation}  
	where $\dot{Q}$ is the rate of Peltier heating for a current ($I$) from Al into Py, $\Pi_{\text{Al(Py)}}$ is the Peltier coefficient of Al (Py), and $R_{\text{th}}$ is the total thermal resistance at the Py/Al junction. In analogy to the standard spin diffusion formalism used to calculate spin resistance $R_{\text{S}}$ \cite{takahashi_spin_2003,maassen_contact-induced_2012}, we implement an analytical heat diffusion model that allows us to calculate $R_{\text{th}}$ \cite{vera-marun_direct_2016,see_supplementary_material_see_????}. Common to both models, such a resistance is dependent on the corresponding conductivity and the characteristic decay length of the corresponding accumulation. For the thermal model, we consider the thermal conductivity $\kappa$ and a thermal transfer length $L_\text{T}$ given by the non-conserved heat current along the metal channel due to the heat flow into the SiO$_2$/Si substrate \cite{vera-marun_direct_2016,see_supplementary_material_see_????,bae_imaging_2010}, which leads to $L_\text{T}\approx900$~nm in the Al channel at 300~K. The total thermal resistance experienced at the injector junction is $R_{\text{th}} \approx 8.8\times10^{5}$~K/W, which is dominated by the higher $\kappa$ of the Al channel. From Eq.~\ref{eq:PeltierTemperature}, the temperature difference at the injector was found to be $\Delta T \approx 1.7$~K, which is in good agreement with the temperature profile of the device area as shown in Fig.~\ref{fig:Simulation}(a) (simulated by 3-dimensional finite element modelling, described later in the text). A non-local Seebeck signal $V_{\text{th}}$ is generated due to $\Delta T$ at a distance $L$ from the injector, given by
	\begin{equation}
		V_{\text{th}} = (S_{\text{Al}}-S_{\text{Py}})\Delta T e^{-L/L_{\text{T}}}.
		\label{eq:SeebeckThermovoltage}
	\end{equation}
	The modelled thermal signal ($V_{\text{th}}/I$) is shown as a function of $L$ in Fig.~\ref{fig:Simulation}(b), together with the experimental baseline resistance ($R_{\text{B}}$). The agreement confirms the thermoelectric origin of the latter, with $R_{\text{B}} \approx V_{\text{th}}/I$. The measured first harmonic response shown in Figs.~\ref{fig:Anisotropy} and \ref{fig:Simulation} is in the linear regime accounting only for the Peltier heating/cooling and therefore excludes Joule heating. Without having used any fitting parameters, our analytical model is accurate within a factor of 2 of the experimentally obtained results. This model disregards lateral heat spreading in the narrow channel and hence serves as an upper estimate of $R_{\text{B}}$ \cite{vera-marun_direct_2016,see_supplementary_material_see_????,bae_imaging_2010}. 
	Furthermore, considering the Thomson-Onsager relation $\Pi= ST$ and a linear temperature dependence of the Seebeck coefficient, we predict a non-linear dependence of $R_{\text{B}}$, which is also substantiated by the measurements in Fig.~\ref{fig:Anisotropy}(b).
	
	We address next our central hypothesis that the anisotropy in the Hanle measurements ($\delta_{\text{anis}}$) emerges via the anisotropy in the thermoelectric coefficients of Py. To account for these magnetothermoelectric effects \cite{wegrowe_anisotropic_2006,mitdank_enhanced_2012}, we relate the isotropic ($R_{\text{B}}$) and the anisotropic ($\delta_{\text{anis}}$) thermoelectric signals, since from Eqs.~\ref{eq:PeltierTemperature}~and~\ref{eq:SeebeckThermovoltage} and the Thomson-Onsager relation, we find that $V_{\text{th}} \propto \Pi_{\text{Py}}.S_{\text{Py}} \propto \Pi_{\text{Py}}^2$. This allows us to explain the ratio $\delta_{\text{anis}}/R_{\text{B}} \approx 2\%$, observed in Fig.~\ref{fig:Anisotropy}d, by considering an anisotropy in the thermoelectric coefficients of Py ($\Pi_{\text{Py}}$, $S_{\text{Py}}$) of approximately 1$\%$. This direct extraction of the anisotropy, $\Delta \Pi_{\text{Py}}/\Pi_{\text{Py}} \approx 1\%$, allows us to successfully model both the channel length ($L$) and temperature ($T$) dependence of the thermoelectric signals, as shown in Fig.~\ref{fig:Simulation}(b-d). 
	
	For completeness, we consider a different anisotropic effect: the modulation in the thermal conductivity of Py, and hence on $R_{\text{th}}$, as a consequence of AMR and the Wiedemann-Franz law. Taking the measured AMR~$= 2$\% at room temperature as an upper limit \cite{kimling_anisotropic_2013}, we obtain an anisotropy which is lower by an order of magnitude than the measured one, and therefore cannot account for the observations. The negligible modulation via this effect is understood by the dominant role of the Al channel (which has no AMR) in determining the total $R_{\text{th}}$.
	
	Finally, an accurate 3-dimensional finite element model (3D-FEM) was developed incorporating the physics of both the anisotropy of the thermoelectric coefficients and of AMR. It is seen in Fig.~\ref{fig:Simulation}(b)-(c) that the 3D-FEM shows a good agreement with the data. A detailed description of the model is included in the supplementary material  \cite{see_supplementary_material_see_????}. Having established the thermal origin of the baseline resistance and the anisotropy, we use this 3D-FEM to explore the asymmetry ($\delta_{\text{asym}}$) observed in the Hanle measurement at 300~K. A finite component of the heat current in the Py bar at the detector junction flowing along the length of the Al channel, combined with the Py magnetization pointing in the out-of-plane direction, generates a transversal voltage along the main axis of the Py bar due to the the anomalous Nernst effect \cite{slachter_anomalous_2011, hu_anomalous_2013}. This transversal voltage gives rise to the asymmetry observed in the Hanle measurements. We successfully account for $\delta_{\text{asym}}$ by considering an anomalous Nernst coefficient of Py equal to 0.06, a factor of two smaller than obtained earlier in Py/Cu spin valves \cite{slachter_anomalous_2011}.
	
	The magnetothermoelectric effects here described are general phenomena in Hanle experiments. Note that the use of tunnel interfaces in previous studies \cite{jedema_electrical_2002-1,tombros_anisotropic_2008,guimaraes_controlling_2014} enhances the spin signal by about 100 times, but from our thermal model that would only amount to an enhancement of the thermoelectric response by a factor of 1. This allows us to understand why the anisotropic signatures have not been identified in previous studies, as the thermoelectric response would only be a modulation of approximately 1\% relative to the spin dependent Hanle signal in those studies. In this work, with transparent contacts and at room temperature, the spin signals are comparable to the thermoelectric effects, making the latter relevant for correct interpretation of the spin-dependent signals.
	
	
	\begin{acknowledgments}
		We thank J.\ G.\ Holstein, H.\ M.\ de Roosz, H.\ Adema and T.\ Schouten for their technical assistance. 
		We acknowledge the financial support of 
		the Zernike Institute for Advanced Materials and 
		the Future and Emerging Technologies (FET) programme within the Seventh Framework Programme for Research of the European Commission, under FET-Open Grant No.~618083 (CNTQC). 
		
	\end{acknowledgments}
	
	
%

\newpage

\appendix

\section{Supplementary Material}

\subsection{Device fabrication}

The Py/Al non-local spin valves (NLSVs) used in this study were prepared by conventional electron beam lithography (EBL), e-beam evaporation and lift-off techniques. A silicon wafer with 300 nm thick thermally oxidized layer on top was used as the substrate. In the first EBL step, the ferromagnets with different widths (70 nm and 90 nm) in order to obtain different coercive magnetic fields, were patterned and followed by the deposition of 20 nm thick Py. In the next step, the 100 nm wide Al channel was patterned and 50 nm thick Al was deposited. In-situ ion-milling, performed just before the deposition of Al, ensured clean, transparent interface between Py/Al. Top Ti/Au contacts were fabricated in the final EBL step. We measured 11 samples with varying injector-detector separation ($L$) prepared in the same batch. The electrical resistivities of the Al and Py wires were measured to be $7.6\times 10^{-8}$ $\Omega.$m ($1.15\times 10^{-7}$ $\Omega.$m) and $3.35\times 10^{-7}$ $\Omega.$m ($4.76\times 10^{-7}$ $\Omega.$m) at 4.2 K (300 K), respectively. The contact resistances of the Py/Al junctions were accurately measured by the 4-probe method where the current was sourced between one end of the Al channel and one end of the ferromagnet wire, while the voltage was measured between the other end of the Al channel and the other end of the same ferromagnet wire, and found to be less than $10^{-15}~\Omega$.m$^2$, thus confirming their transparent nature \cite{villamor_effect_2015}.

\subsection{Anisotropic Magnetoresistance measurements}

\begin{figure*}[tbp]
	\includegraphics[angle=0, trim=0mm 0mm 0mm 0mm, width=\textwidth]{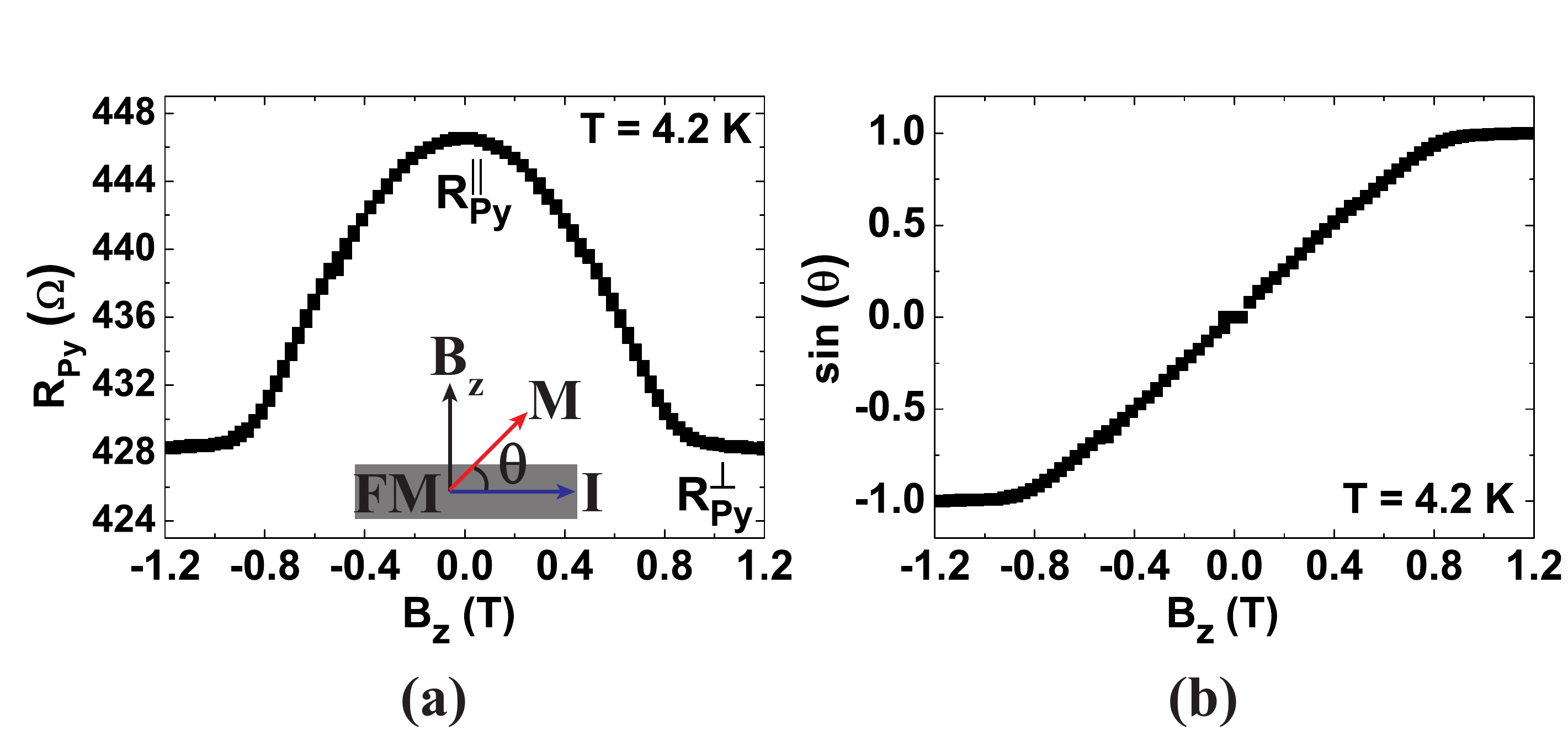}
	\caption{\textbf{AMR measurement of Py} \textbf{(a)} Anisotropic magnetoresistance curve for a Py wire. The inset shows a schematic for the out-of-plane rotation of the Py magnetization ($M$) due to $B_{\text{z}}$. \textbf{(b)} Sine of the angle ($\theta$) between $M$ and the current ($I$) in Py, extracted from the measurement in \textbf{(a)}.}
	\label{FigS0}
\end{figure*}

Anisotropic magnetoresistance (AMR) measurements were carried out on the Py wires to probe the variation of their electrical resistance ($R_\text{Py}$, measured in four-terminal configuration) with the out-of-plane magnetic field ($B_\text{z}$), as shown in Fig.~\ref{FigS0}(a). From this AMR curve, we can extract the angle ($\theta$) between the magnetization and the current ($I$) direction as a function of $B_{\text{z}}$ by using the expression \cite{rijks_semiclassical_1995,jedema_electrical_2002-1}, 
$R_{\text{Py}}(B_{\text{z}})=R_{\text{Py}}^{\perp}+[R_{\text{Py}}^{\parallel}-R_{\text{Py}}^{\perp}]\cos^{2}\theta(B_{\text{z}})$, 
where $R_{\text{Py}}^{\parallel}$ ($R_{\text{Py}}^{\perp}$) is the resistance of the Py wire when its magnetization is oriented in-plane (out-of-plane), parallel (perpendicular) to the direction of $I$. 
The sine of $\theta$, plotted in Fig.~\ref{FigS0}(b), shows that for $|B_{\text{z}}|\geq 0.9$~T the magnetization of Py is completely aligned with $B_{\text{z}}$. 

\subsection{Hanle data fitting}
The Hanle data was fitted with an analytical expression taking into consideration the transparent nature of the contacts \cite{fukuma_giant_2011,villamor_effect_2015} and the out-of-plane rotation of the Py magnetization \cite{jedema_electrical_2002-1}:

\begin{widetext}
\begin{equation}
R_{\text{NL}}^{\text{P(AP)}}(B_\text{z},\theta) = \pm R_{\text{NL}}(B_\text{z})\cos^2\theta + |R_{\text{NL}}(B_\text{z}=0)|\sin^2\theta,
\label{eq:Hanle Fit}
\end{equation}
\end{widetext}
where, $\theta$ is the angle between the Py magnetization and the plane of the Py film. The expression for $R_{\text{NL}}(B_\text{z})$ is obtained by solving the Bloch equation considering spin precession, diffusion and relaxation for transparent contacts, explicitly mentioned elsewhere \cite{fukuma_giant_2011,villamor_effect_2015}. Using Eq.~\ref{eq:Hanle Fit} to fit the Hanle data obtained at low temperatures (4.2~K and 80~K), where the anisotropy in the Hanle line-shape is absent, we extracted values for the spin relaxation length in Al ($\lambda_{\text{Al}}$) which are in close agreement with those obtained from the spin-valve measurements (see main text). However, the Hanle data with the anisotropic line-shapes ($T>150$~K) cannot be fitted by using Eq.~\ref{eq:Hanle Fit}. After the origin of this anisotropy was established, we modified Eq.~\ref{eq:Hanle Fit} by adding a term to account for the anisotropic Hanle line-shape, as shown below:

\begin{widetext}
\begin{equation}
R_{\text{NL}}^{\text{P(AP)}}(B_\text{z},\theta) = \pm R_{\text{NL}}(B_\text{z})\cos^2\theta + |R_{\text{NL}}(B_\text{z}=0)|\sin^2\theta + R_{\text{B}}(1+\frac{\delta_{\text{anis}}}{R_{\text{B}}}\sin^2\theta) ,
\label{eq:Anis Hanle Fit}
\end{equation}
\end{widetext}
with $R_{\text{B}}$ and $\delta_{\text{anis}}$ the baseline resistance and the anisotropy, respectively. The extra term on the R.H.S. of Eq.~\ref{eq:Anis Hanle Fit} includes the baseline resistance and its modulation due to the anisotropic magnetothermoelectric effects described in the main text.

\subsection{Analytical Heat Diffusion Model}

We use a simplistic thermal transport model \cite{vera-marun_direct_2016} for giving us a physical insight into the origin of the anisotropy in the Hanle measurements ($\delta_{\text{anis}}$) and attribute it to the anisotropic modulation of the baseline resistance ($R_{\text{B}}$) via the anisotropic magnetothermoelectric effects as discussed in the main text. This model considers one-dimensional diffusive heat transport in the metal channels (Al and Py) with a point (Peltier) heat source at the Py/Al injector junction. The heat flow in the metal channels is non-conserved since it flows into the SiO$_2$/Si substrate acting as a heat reservoir. The thermal transfer length ($L_{\text{T,~M}}$) in a metallic channel (M = Al, Py), described as the average distance over which the heat flows in that channel, is given by \cite{bae_imaging_2010}
\begin{equation}
L_{\text{T,~M}} = \sqrt{\kappa_{\text{M}}t_{\text{M}}t_{\text{ox}}/\kappa_{\text{ox}}} ,
\label{eq:Thermal Transfer Length}
\end{equation}    

with $\kappa_{\text{M}}$ ($\kappa_{\text{ox}}$) and $t_{\text{M}}$ ($\kappa_{\text{ox}}$) the thermal conductivity and the thickness of the metallic channel (SiO$_2$), respectively. Using the material parameters listed in Table~\ref{table1}, we calculated the thermal transfer lengths in Al and Py to be 870~nm and 270~nm, respectively, at 300~K. These lengths being larger than the dimensions of the Py/Al junctions, support our model assumption of treating the Py/Al junctions as point contacts. In analogy to the spin resistance described in diffusive spin transport models \cite{takahashi_spin_2003}, the thermal resistance $R_{\text{th,~M}}$ of the metal channel can now be calculated as

\begin{equation}
R_{\text{th,~M}} = \frac{L_{\text{T,~M}}}{2\kappa_{\text{M}}w_{\text{M}}t_{\text{M}}} ,
\label{eq:Thermal Resistance}
\end{equation}

where, $w_{\text{M}}$ is the width of the metal channel. Using Eq.~\ref{eq:Thermal Resistance}, we calculated $R_{\text{th,~Al}}\approx1.1\times10^6$~K/W and $R_{\text{th,~Py}}\approx4.3\times10^6$~K/W. The total thermal resistance $R_{\text{th}}$ at the Peltier junction is then expressed as $R_{\text{th}}=R_{\text{th,~Al}} \parallel R_{\text{th,~Py}}\approx 8.8\times10^5$~K/W, which is clearly dominated by the Al channel with a higher thermal conductivity. This simplistic analytical model does not use any fitting parameters and serves as an upper estimate of the thermal resistance since it does not take into consideration the lateral heat spreading in the metallic channels and neglects the finite widths of the Py/Al junctions, treating them as point contacts.

\subsection{Three-dimensional finite element simulation (3D-FEM)}

Using a 3D-FEM, we calculated the spin signal $R_{\text{S}}=R_{\text{NL}}^{\text{P}}-R_{\text{NL}}^{\text{AP}}$, the baseline resistance $R_{\text{B}}=(R_{\text{NL}}^{\text{P}}+R_{\text{NL}}^{\text{AP}})/2$ and the observed anisotropy $\delta_{\text{anis}}$ due to magnetothermoelectric effects for devices with transparent contact properties. 
The gradients in the spin-dependent electrochemical potential $\vec{\nabla} V_{\uparrow,\downarrow}$ and temperature $\vec{\nabla} T$ are related to the respective charge $\vec{J}_{\uparrow,\downarrow}$ and heat current $\vec{Q}$ as\cite{slachter_anomalous_2011}
\vspace{-0.3cm}


\begin{subequations}\label{eqS2}
	\begin{eqnarray}
	\vec{J}_{\uparrow,\downarrow}=-\sigma_{\uparrow,\downarrow}\vec{\nabla}V_{\uparrow,\downarrow}-\sigma_{\uparrow,\downarrow}S_{\uparrow,\downarrow}\vec{\nabla} T\\
	\vec{Q}=-\sigma_{\uparrow,\downarrow}\Pi_{\uparrow,\downarrow}\vec{\nabla}V_{\uparrow,\downarrow}-\kappa\vec{\nabla}T
	\end{eqnarray}
\end{subequations}

\begin{figure*}
	\includegraphics[angle=0, trim=0mm 0mm 0mm 0mm, width=\textwidth]{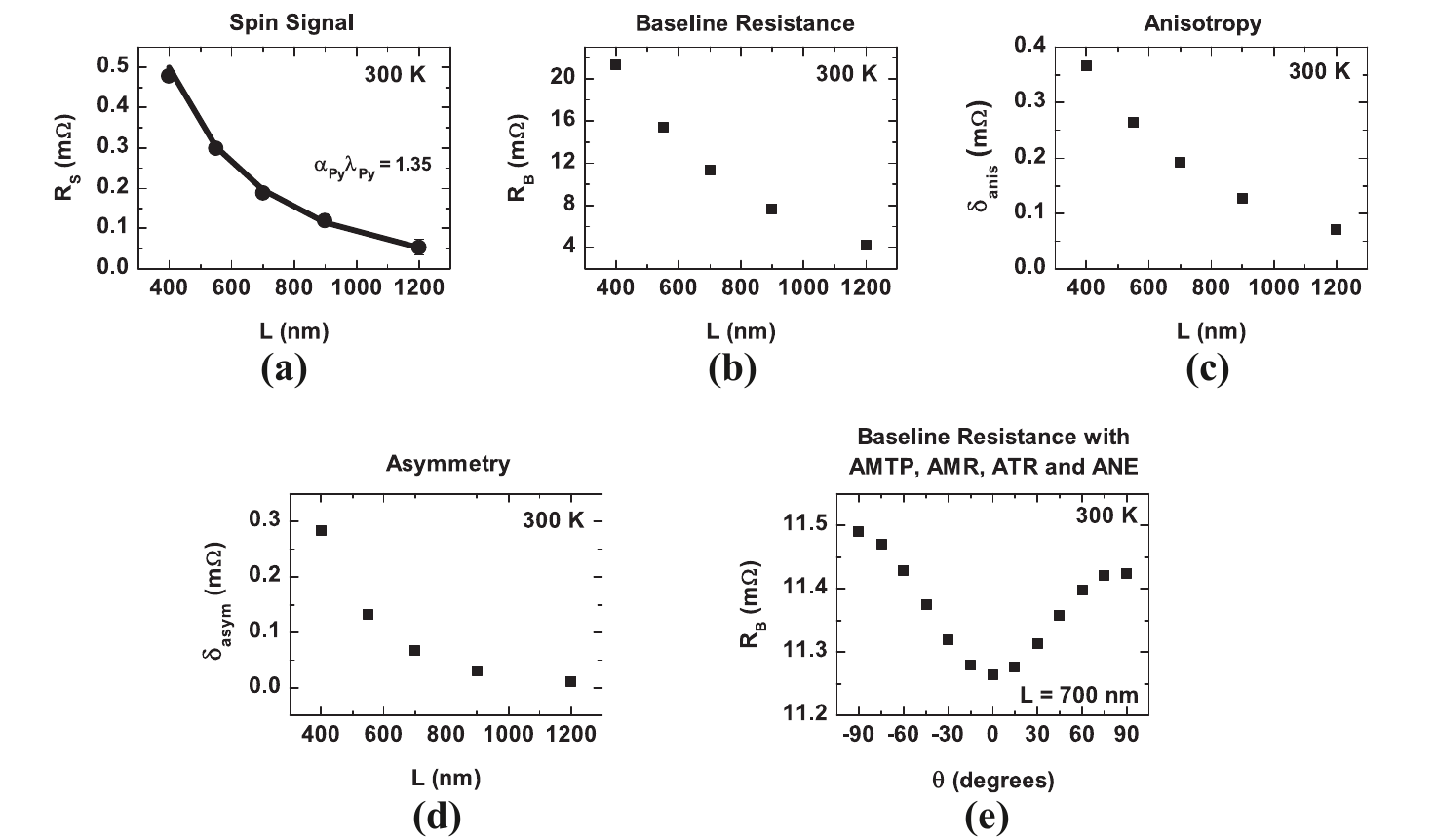}
	\caption{\textbf{Three-dimensional finite element modelling (3D-FEM).} \textbf{(a)} Numerical simulation results (solid line) showing the injector-detector distance ($L$) dependence of $R_\text{S}$ at 300~K plotted along with the measured data (black circles). The product $\alpha_{\text{Py}}\lambda_{\text{Py}}$, which is the adjustable fitting parameter, is indicated. The corresponding $R_\text{B}$ calculated from the 3D-FEM for different $L$ values at 300~K is shown in \textbf{(b)} along with its anisotropy $\delta_{\text{anis}}$ in \textbf{(c)} and asymmetry $\delta_{\text{anis}}$ in \textbf{(d)}. Dependence of $R_{\text{B}}$ on the angle ($\theta$) between the magnetization direction of Py and the \textit{y}-axis is plotted in \textbf{(e)} at 300K and for $L=700$~nm, with AMTP = 1\%, AMR = ATR = 2\% and anomalous Nernst coefficient $R_\text{N}$ = 0.06. The 3D-FEM model more accurately describes the observed anisotropy and asymmetry in $R_\text{B}$ as well as the measured spin signals for a range of temperatures and distances.}
	\label{FigS1}
\end{figure*}

\noindent where $\sigma_{\uparrow,\downarrow}=\frac{\sigma}{2}(1\pm P_\sigma)$ is the spin-dependent electrical conductivity described in terms of the spin-polarization $P_\sigma=(\sigma_\uparrow-\sigma_\downarrow)/\sigma$ of the electrical conductivity $\sigma$ and the spin-up $\sigma_\uparrow$ and spin-down $\sigma_\downarrow$ conductivities,  $\Pi_{\uparrow,\downarrow}=S_{\uparrow,\downarrow}T_0$ is the spin-dependent Peltier coefficient related to the spin-dependent Seebeck coefficient $S_{\uparrow,\downarrow}$ via the Thomson-Onsager relation and $\kappa$ is the thermal conductivity. In magnetic metals, these transport coefficients are tensors that encompass anisotropy.

To include the anisotropic magnetoresistance (AMR), anisotropic thermoresistance (ATR), anisotropic magnetothermopower (AMTP) and the anomalous Nernst effect (ANE) we use anisotropic transport coefficients \cite{slachter_anomalous_2011} that depend on the relative orientation of the unit magnetization vector $\hat m$ pointing in the direction of the magnetization of the ferromagnet with that of $\vec J$ and/or $\vec Q$. For $\hat m$ making angles $\theta$ with the x- and $\phi$ with the z-axis, the expression for the anisotropic transport coefficients are:

\begin{table*}
	\caption{Material parameters for the 1D analytical and 3D numerical simulations. The electrical conductivity $\sigma$ is obtained from four-terminal resistance measurements and $\kappa$ is calculated from the measured $\sigma$ using the Wiedemann-Franz (WF) relation $\kappa=L_0\sigma T_0$, where $L_0=2.44\times 10^{-8}$~W$\Omega$K$^{-2}$ and $T_0$ are the Lorenz number and the reference temperature, respectively. The temperature ($T$) dependent $S$ of Al and Py are calculated by interpolating the known values at room temperature via the expression\cite{bakker_interplay_2010} $S=S_0\cdot T/T_0$. This approximation, following from Mott's relation for S, does not however include magnon and phonon drag contributions.}
	\begin{ruledtabular}
		\begin{tabular}{l c c c c}
			Material [thickness] & $\sigma$ [$10^6$ S/m]& $S$ [$\mu$V/K] & $\kappa$ [W/(mK)] & $\lambda_s$ [nm] \\ \hline
			
			Al (50 nm) & 8.73 & -1.5 & 64 & 383 \\
			Ni$_{80}$Fe$_{20}$ (20 nm) & 2.1 & -18 & 15.5 & 5  \\
			Au (120 nm)  & 27  & 1.6 & 80 & - \\
			SiO$_2$ (300 nm) & 1e-18 & 1e-12 & 1.2 & -\\
			Si (0.5 mm) & 0.01  & -100 & 80 & 1000
			
		\end{tabular}
	\end{ruledtabular}
	\label{table1}
\end{table*}

\begin{widetext}
\begin{subequations}\label{eq:2}
	\begin{eqnarray}
	\sigma_{ij}=\sigma_{\perp}\left(\delta_{ij}-R_{\text{AMR}} \hat m_{i} \hat m_{j} \right),\label{eq:2a}\\
	S_{ij}=S_{\perp}\left(\delta_{ij}-S_{\text{AMTP}} \hat m_{i} \hat m_{j}\right)+S\left(\delta_{ij}-R_{\text{ANE}} \sum_{k} \varepsilon_{ijk} \hat{m}_{k} \right),\label{eq:2b}\\
	\kappa_{ij}=\kappa_{\perp}\left(\delta_{ij}-R_{\text{ATMR}} \hat m_{i} \hat m_{j} \right)~\text{and}\label{eq:2c}\\
	\Pi_{ij}=S_{ij}T_0.
	\end{eqnarray}
\end{subequations}
\end{widetext}
	
\noindent Here $i,j=x, y, z$ define the components of $\hat m$ along the principal axes, $\delta_{ij}$ and $\varepsilon_{ijk}$ are the Kronecker delta and Levi-Civita symbols, respectively. Eq.~\ref{eq:2a} describes the anisotropic magnetoresistance effect with $R_{\text{AMR}}=(R_\parallel-R_\perp)/R_\parallel$ being the experimentally determined AMR ratio. Using the Wiedemann-Franz relation $\kappa\propto\sigma$, the anisotropic thermoresistance (ATR) ratio $R_{\text{ATMR}}$ in Eq.~\ref{eq:2c} is set to $R_{\text{AMR}}$ in Eq.~\ref{eq:2a}. While the first term in Eq.~\ref{eq:2b} represents the anisotropic magneto thermopower (AMTP), the second term describes the anomalous-Nernst effect \cite{slachter_anomalous_2011}. Thermal transport through the 300 nm thick SiO$_2$ to the bottom of the Si substrate, that is set as a thermal sink, is also included. We do not take the whole thickness $t_{\text{sub}}=0.5$ mm of the Si/SiO$_2$ substrate but model the Si substrate as a $20~\mu$m cube with a thermal conductivity  $\kappa_{\text{Si}}=80$ Wm$^{-1}$K$^{-1}$. The input material parameters for the 3D-FEM are shown in Table~\ref{table1}.

Our simulation procedure is as follows. First, we obtain $P_\sigma$ by matching the model to the measured spin signals for the injector-detector distance of 400 nm. Here setting $\lambda_{\text{Al}}$ to the spin-diffusion length values obtained from the 1D spin-diffusion model (see main text) and $\lambda_{\text{Py}}=5$ nm, at room temperature, we obtain $P_\sigma=0.28$, in good agreement with earlier reports in similar spin transport measurement configurations\cite{jedema_electrical_2001,villamor_effect_2015}. The calculated baseline resistance $R_\text{B}$ of 22 m$\Omega$ is lower than the measured value of 32 m$\Omega$, which indicates to the fact that the $\kappa_{\text{Si}}$ used in the simulation might be larger than in our devices. Next we keep both $\lambda_{\text{Py}}$ and $P_\sigma$ constant and calculate, for instance, the distance and angle dependence of $R_\text{S}$, $R_\text{B}$ and $\delta_{\text{anis}}$. To accurately describe the observed anisotropy and asymmetry in $R_\text{B}$ we use the experimentally obtained anisotropic magnetoresistance ratio $R_{\text{AMR}}=0.02$. Following WF relation, the anisotropic magnetothermal resistance ratio $R_{\text{ATMR}}$ is set equal to $R_{\text{AMR}}=0.02$. By varying $S_{\text{AMTP}}$, that describe the anisotropy in $R_\text{B}$, and the $R_{\text{ANE}}$, that quantifies asymmetry due to the anomalous Nernst effect, we can fully quantify the observed anisotropy in $R_\text{B}$ at room temperature for all measured distances. We note that both the anisotropy and asymmetry in the baseline resistance are not caused by the conventional AMR effect, but instead by the anisotropy in the Seebeck (Peltier) coefficients. In Fig.~\ref{FigS1}, we present some results from the 3D-FEM simulation.

\subsection{Additional Experiments: Higher harmonic measurement, Current Density Dependence and Check for Inverted Hanle}

\begin{figure*}
	\includegraphics[angle=0, trim=0mm 0mm 0mm 0mm, width=\textwidth]{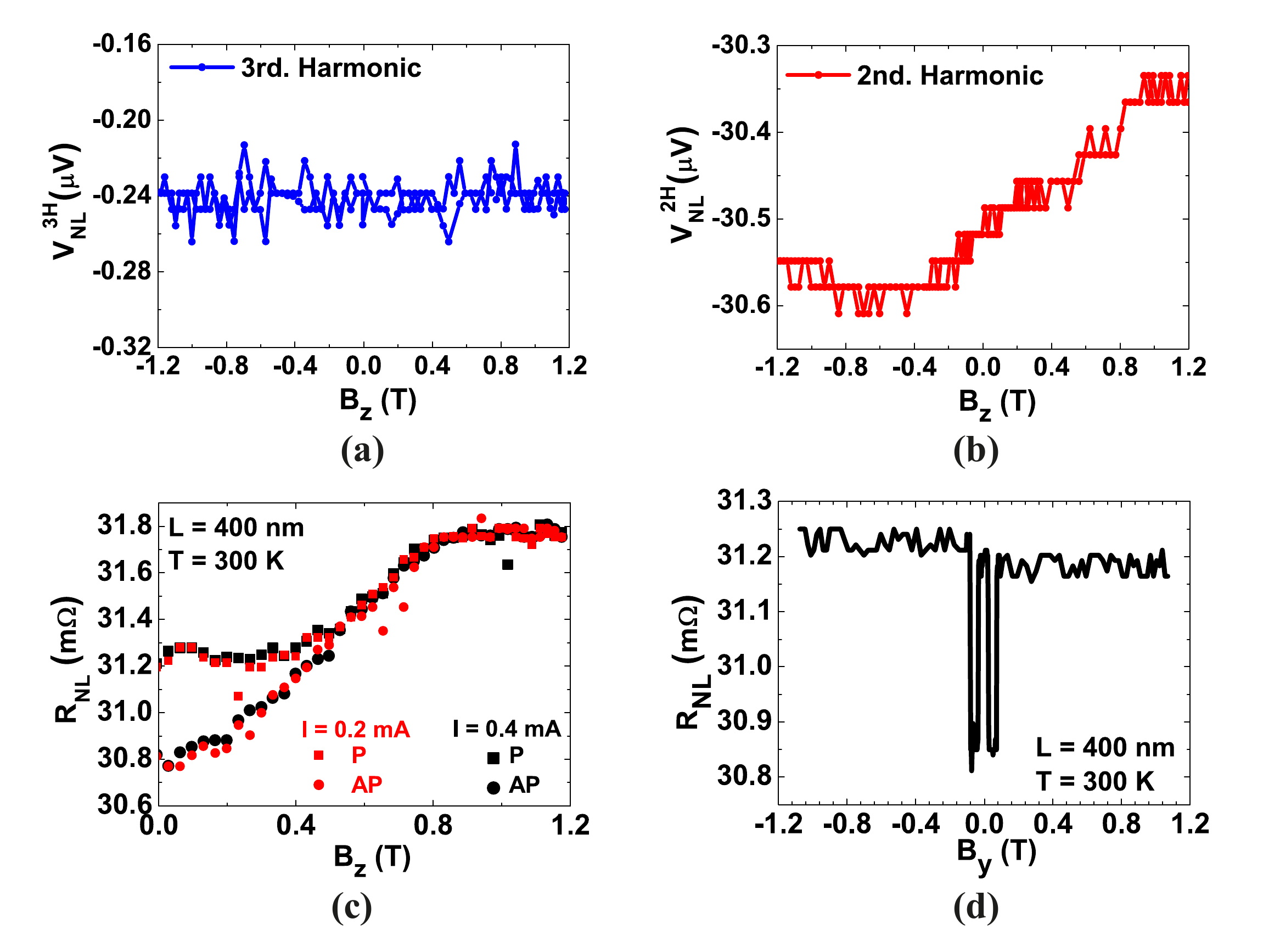}
	\caption{\textbf{Additional measurements to rule out spurious effects.} \textbf{(a)} The third harmonic, and \textbf{(b)} the second harmonic responses at $T=300$~K with injection current $I=0.4$~mA for a NLSV with $L=400$~nm. The featureless third harmonic response excludes the presence of non-linear effects in the first harmonic signal of the Hanle measurements. The modulation in the second harmonic response due to Joule heating is almost negligible compared to the Peltier effect in the first harmonic. \textbf{(c)} Hanle measurements (first harmonic response) at two different current densities $I$, 0.2~mA (red symbols) and 0.4~mA (black symbols) on the same NLSV with $L=400$~nm and at $T=300$~K. The corresponding non-local signals ($R_{\text{NL}} = V_{\text{NL}}/I$) are the same and confirm that the measurements are carried out in the linear response regime. \textbf{(d)} The absence of inverted Hanle demonstrated by carrying out the spin-valve measurements using high in-plane magnetic fields ($B_{\text{z}}$). This rules out the role of local magnetostatic fields arising from interfacial roughness.}
	\label{FigS2}
\end{figure*}

Several checks were performed to rule out spurious effects contributing to the anisotropy in the Hanle line-shape. First, we perform higher harmonic detection using the lock-in technique and measure the third harmonic and second harmonic responses of the Hanle measurements at $T=300$~K with an alternating current (a.c.) $I$ of 0.4~mA at frequency $f<15$~Hz, as shown in Figs.~\ref{FigS2}(a) and (b) respectively. The featureless third harmonic response, occurring at $3f$, serves as a clear proof of the absence of non-linear contributions in the first harmonic signal above the noise level in our measurements. The second harmonic response, occurring at $2f$, represents the contribution from Joule heating and shows a negligible modulation as compared to the linear Peltier effect detected in the first harmonic signal. Next, we repeat the Hanle measurements at two different current densities ($I=0.2$~mA and 0.4~mA) and show the first harmonic signal in Fig.~\ref{FigS2}(c). The non-local response ($R_{\text{NL}} = V_{\text{NL}}/I$) in both cases are the same and this further confirms that the measurements are carried out in the linear response regime.

We also probed the effect local magnetostatic fields caused by interfacial roughness \cite{dash_spin_2011} and magnetic impurities \cite{mihajlovic_magnetic-field_2011} that might result in the anisotropic Hanle line-shape. The experimental signature of the presence of these local magnetostatic fields is the inverted Hanle signal. In order to detect it, we applied high magnetic fields ($\pm 1.1$~T) parallel to the interface ($B_{\text{y}}$) in the spin-valve measurement configuration. The absence of any inverted Hanle line-shape (see Fig.~\ref{FigS2}(d)) even at these high magnetic fields confirms the absence of local magnetostatic fields due to interfacial roughness.

\begin{figure*}
	\includegraphics[angle=0, trim=0mm 0mm 0mm 0mm, width=\textwidth]{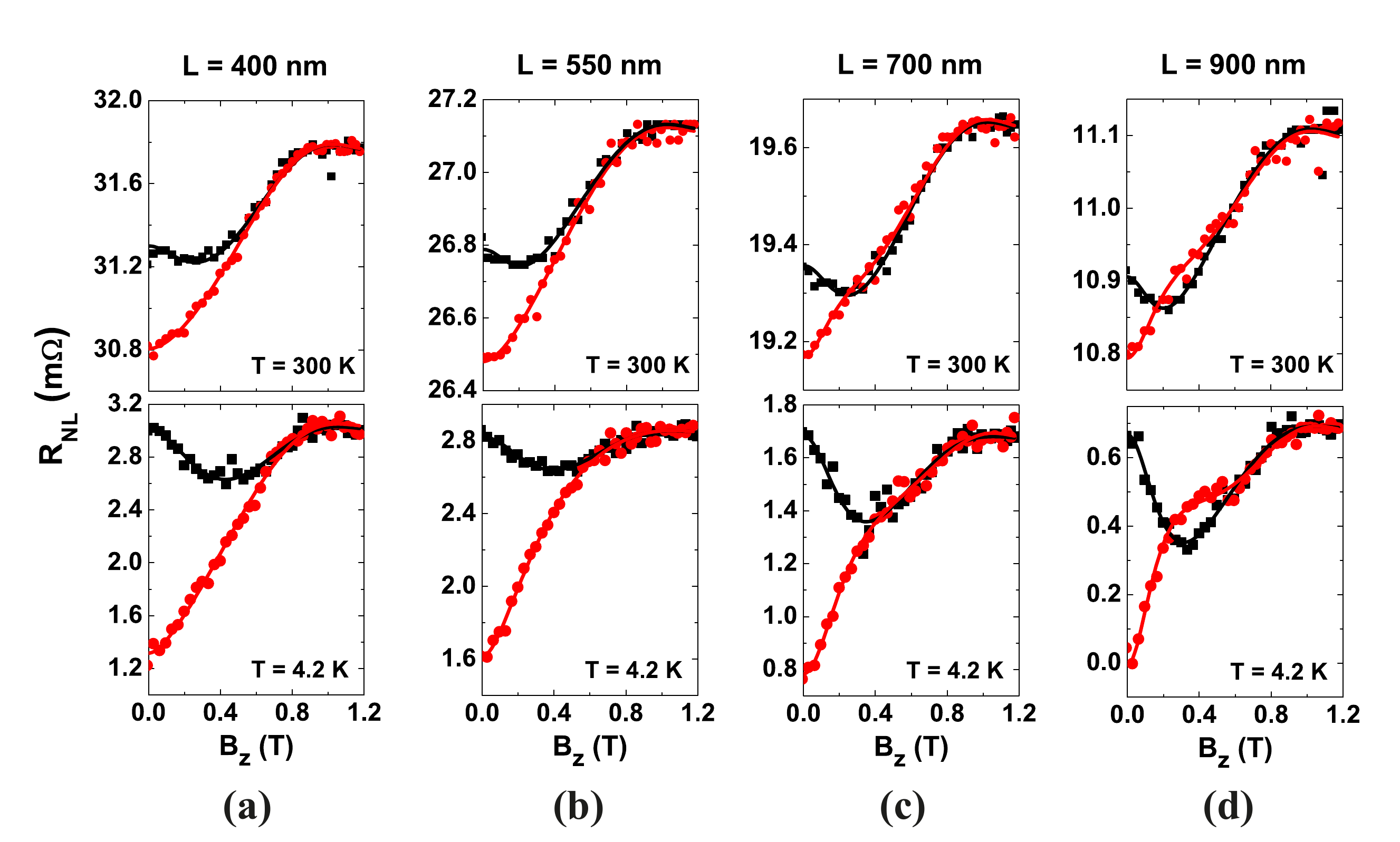}
	\caption{\textbf{Hanle measurements for different spacings between the ferromagnets.} Hanle data at $T=300$~K (top panels) and  $T=4.2$~K (bottom panels) for NLSVs with different injector-detector spacings: \textbf{(a)} $L=400$~nm, \textbf{(b)} $L=550$~nm, \textbf{(c)} $L=900$~nm and \textbf{(d)} $L=900$~nm.}
	\label{FigS3}
\end{figure*}

\end{document}